\documentclass[pre,aps,showpacs,amsmath,amssymb,preprint,endfloats,letterpaper]{revtex4}

\usepackage{graphicx}
\usepackage{mathrsfs}
\usepackage{wasysym}
\usepackage{textcomp}
\usepackage{pifont}
\usepackage{stmaryrd}
\usepackage{amssymb}

\begin{document}

\title{Laser-to-proton energy transfer efficiency in laser-plasma interactions}
\author{E. Fourkal}
\author{I. Velchev}
\author{C-M. Ma}

\affiliation{Department of Radiation Physics, Laser Laboratory, Fox Chase Cancer Center,
Philadelphia, PA 19111, U.S.A.}




\begin {abstract}
It is shown that the energy of protons accelerated in laser-matter interaction experiments may be significantly increased through the process of splitting the incoming laser pulse into multiple interaction stages of equal intensity. From a thermodynamic point of view, the splitting procedure can be viewed as an effective way of increasing the efficiency of energy transfer from the laser light to protons, which peaks for processes having the least amount of entropy gain. It is predicted that it should be possible to achieve $\apprge 100\%$ increase in the energy efficiency in a six-stage laser proton accelerator compared to a single laser-target interaction scheme.           
\end {abstract}


\date{\today}

\maketitle

\newpage



\section{Introduction}
Ion acceleration by high-power lasers has attracted significant attention in recent years from the scientific community due to its potential applications in different branches of physics and technology. The physical characteristics of accelerated protons such as high collimation and high particle flux make them very attractive for possible applications in controlled nuclear fusion~\cite{roth01,bychenkov01}, material science~\cite{boody96}, and hadron therapy~\cite{khoroshkov02,fourkal02}. 

The physical processes responsible for ion acceleration during laser-matter interaction are understood on a qualitative level. The initially proposed theoretical model for ion acceleration at the back surface of the target heavily relied on the notion of quasi-neutral plasma expansion into vacuum~\cite{gurevich66,kovalev03}. In this model the accelerating
electric field is generated due to space-charge separation in a narrow layer at the front of the expanding plasma cloud, which is assumed to be neutral. In the interaction of ultrashort and ultra-intense laser pulse with a solid structure, the assumption of quasi-neutrality must be abandoned. The results of computer simulations~\cite{bulanov00,sentoku00} suggest that the interaction of petawatt laser pulses with plasma foils leads to the formation of extended regions where plasma quasi-neutrality is violated, a factor that has to be taken into account when considering ion acceleration by ultra-intense pulses ($I \sim 10^{21}$ W/cm$^2$). According to the model, the incoming laser pulse quickly ionizes the target pushing some of the electrons out of it through the action of the ponderomotive force. A strong electrostatic field ($\sim$ TV/m) is set up between the expanding electrons and the target that field ionizes the thin hydrogen-rich layer present at its back surface. Subsequently, the protons are accelerated in this electrostatic field~\cite{bulanov02,fourkal05}. For thicker targets ($\ge$ 2 $\mu$m) a shock wave acceleration mechanism has also been proposed\cite{silva04} in which a laser acts as a piston driving a flow of ions into the target and launching an electrostatic shock at the front of the target with high Mach number M=$v_{\textnormal{shock}}$/c$\simeq$ 0.2-0.3. Protons, reflected off the shock front may get accelerated to velocities up to $v_{\textnormal{ions}}=2v_{\textnormal{shock}}$. 

In a recent multi-parametric particle-in-cell (PIC) simulation study of interaction between a clean (no prepulse present) high-power laser pulse and thin double-layer target~\cite{esirkepov06} the authors have mapped maximum proton energy regions as functions of target electron density and its thickness as well as laser pulse length for different laser intensities and spot sizes. According to the scaling laws obtained by the authors, in order to accelerate protons to the energy range of a few hundred MeV (e.g. required for hadron therapy applications where protons with energy 250 MeV can reach any disease site throughout the patient's body), one needs to pump a few hundred joules of energy or equivalently several tens of petawatt of power (for laser pulse duration $L_p\sim$100 fs) into the laser pulse, provided that optimal conditions for the laser/target parameters are met. Currently available lasers, specifically compact table-top systems, operating in the sub-picosecond regime provide energy on the order of $\mathscr{E}_l\sim$10 J. According to the scaling laws obtained in Ref.~\cite{esirkepov06} this energy is insufficient to accelerate protons to the required therapeutic energy range of 200-250 MeV. Therefore, it would be interesting to investigate whether there is a way to increase the maximum proton energy or equivalently the efficiency of energy transfer from the laser pulse into accelerated protons without increasing the energy of the laser pulse.

In a recent work~\cite{velchev07}, we have shown that in a double layer target system, the acceleration conditions for protons are far from the optimal due to the fact that protons are expelled from the back surface of the substrate before the maximum electric field is established and as a result experience reduced acceleration potential. It has been shown that by modifying the proton dynamics (through splitting the pulse into two interaction stages), higher final proton energies can be achieved. Up to 30 $\%$ increase in the final proton energy as compared to a single interaction stage was predicted through the double splitting procedure. The natural extension of the idea presented above would be to find out whether the energy transfer efficiency from the laser pulse to protons can be further improved. An obvious question that one may ask is: if a single interaction splitting leads to $\sim$ 30 $\%$ increase in the proton average energy, then how would the introduction of $n$ interaction stages influence the final proton energy? 

In this work we will show that splitting a single interaction scheme into $n$ stages gradually increases the energy transfered from the laser pulse to a proton beam with each additional splitting, thus increasing the final energy of the proton beam. We will offer a thermodynamic analogy to this effect, which helps elucidate the underlying physics. In fact, the problem at hand is related to the question of what the most efficient way of transferring the energy from a hot object (laser) to a cold (protons) one is. It is a known fact that any energy exchange process between the constituents of a \emph{closed} system is accompanied by an increase in the total entropy of the system, no matter what the interaction is (the second law of thermodynamics). The magnitude of the entropy increase however depends on the manner in which the energy exchange occurs and smaller increment in the entropy yields higher energy transfer efficiency~\cite{lifshits}. According to classical thermodynamics, a process occurring near equilibrium is the most efficient in terms of generating the maximum amount of work and the least amount of entropy. A reversible thermodynamic process merely needs to be slow enough for all thermally-interacting constituents to equilibrate. This means that the high efficiency of energy conversion can be attained, provided that the relaxation time is much shorter than the characteristic time of the process. We show that the most efficient heat exchange occurs when the cold and hot objects are split into $n$ equal pieces and each individual hot piece is put into thermal contact with each individual cold piece (without mixing them) in a sequential manner (see Appendix). In the end, initially hot/cold pieces are put back together to form a new cold/hot object correspondingly. As the number of splits increases, the entropy change $\Delta S$ for the whole process decreases and in the limit $n\shortrightarrow \infty$, $\Delta S \shortrightarrow 0$~\cite{mishchenko}. In this \emph{asymptotic} case the initially cold object becomes hot (with temperature equal to the initial temperature of the hot object) and initially hot object becomes cold (with temperature equal to the initial temperature of the cold object) and the perfect (completely reversible) heat exchange process is achieved. It should be noted that the introduced heat exchange problem does not violate the second law of thermodynamics, since the energy always flows from the hot object to the cold and increased heat transfer efficiency is achieved through the reduction of the amount of energy exchanged in each individual interaction step, which ultimately reduces the overall energy "wasted" during the heat flow. Therefore, it is reasonable to conjecture that when the laser pulse is split into $n$ sub-pulses of equal intensity $I_0/n$ that are made to interact with $n$ targets (laser-target multi-system constitutes a closed system), the energy transfer efficiency (kinetic energy of the accelerated protons) increases due to a decrease in the total entropy gain for the closed system (laser light, all particles and fields). Just as in the case with hot/cold reservoirs, the splitting procedure is an effective way of reducing the energy wasted during the laser-target interaction, thus increasing the energy transferred from the laser pulse to protons (adiabatic acceleration). 

Bringing the entropy argument into the description of laser-mater interaction physics (essentially nonequilibrium process) and giving it a prominent role comes naturally after a long string of successful applications of the \emph{minimum entropy production  principle}~\cite{prigogine47,robertson69,klein54} to various systems including plasmas. The qualitative analogy we offer is a further demonstration of the far reaching consequences of this principle that may eventually be used in a quantitative analysis of the physics of laser-matter interactions.

\section{Multi-stage proton acceleration in 2D particle-in-cell simulations and 3D model}                                        
\begin{figure}
\centerline{\includegraphics[width=1.\columnwidth]{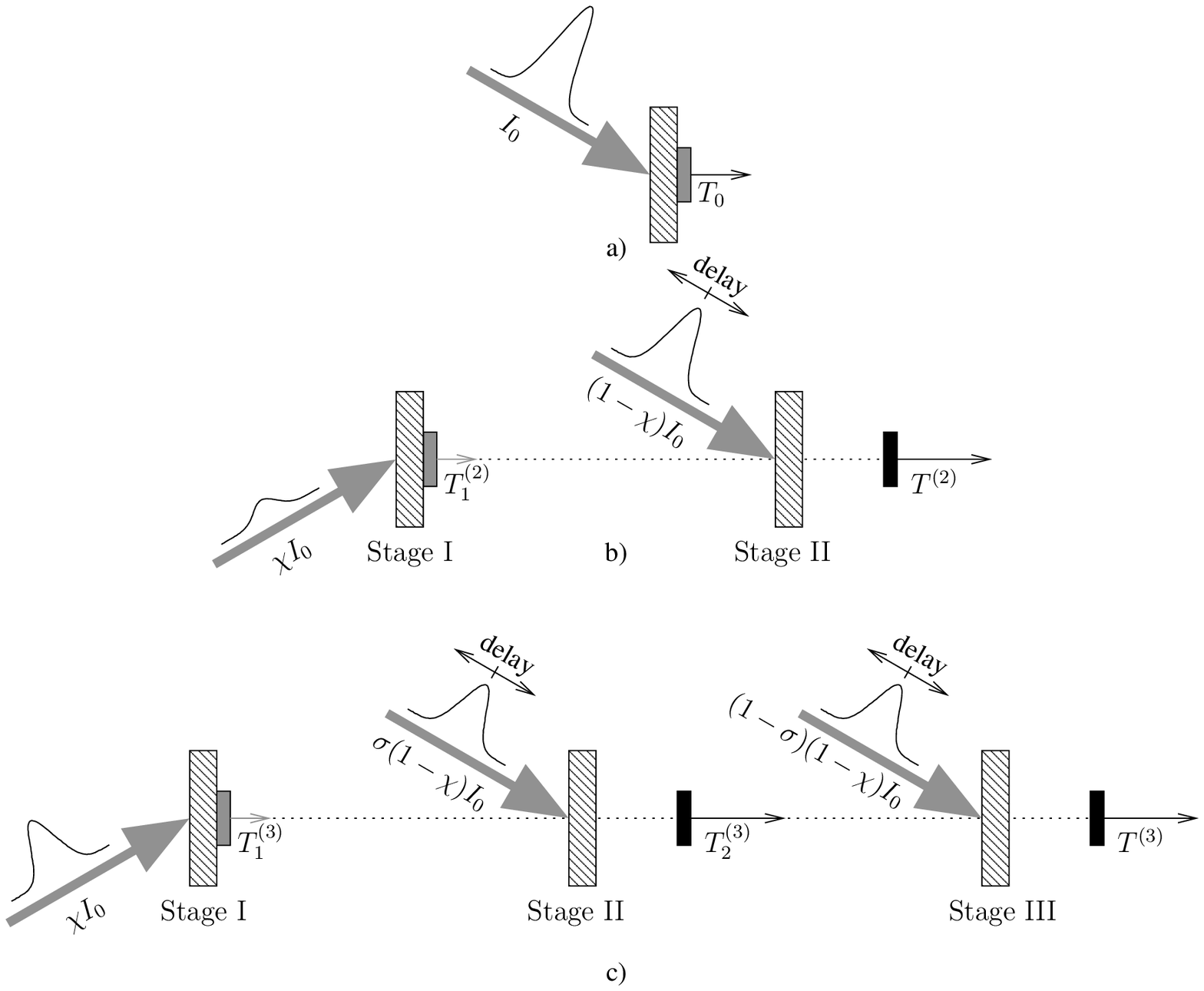}}
\caption{ 
Proposed experimental setup for increased proton energy; a) --
conventional double-layer target geometry; b) -- two-stage proton
acceleration; c)--three-stage proton acceleration}\label{fig1}
\end{figure}

In order to study how the introduction of multiple interaction stages influences the final proton energy, we have used 2D PIC simulations~\cite{tajima,langdon} to model the interaction between the laser pulse and several targets. The calculations were performed in a $2048\times 1024$ simulation box with a grid size $\Delta=0.04$ $\mu$m and total number of simulated quasi-particles $2.6 \times 10^6$. Periodic boundary conditions for particles and electromagnetic fields have been used. The initial conditions were chosen to correspond to realistic experimental parameters, where linearly polarized ($p$ polarization) relativistically intense ($I_0$=1.92$\times$10$^{21}$ W/cm$^2$, $\lambda$= 800 nm), ultrashort (L$_p$ $\approx$ 30 fs) Gaussian laser pulse (focal spot size $D$=3.2 $\mu$m at FWHM) is normally incident in a Cu target of thickness 400 nm and transverse dimension 5 $\mu$m. The electron density as well as the ion charge state in the target are n$_e$=3.2$\times$10$^{22}$ cm$^{-3}$ and $Z_i$=4 correspondingly. A 200 nm thick hydrogen-rich layer (n$_e$=6$\times$10$^{19}$ cm$^{-3}$) with transverse dimension of 2.5 $\mu$m is initially located at the back surface of the target. Two and three interaction stages have been simulated and the final proton energy (averaged over all protons) has been compared to that obtained in a single interaction scheme. The schematic diagram of multi-stage interaction setup is shown in figure~\ref{fig1}. In the multiple interaction scheme, the laser pulse of intensity $I_0$ is split into $n$ sub-pulses of equal intensity $I_0/n$ that are made to interact with $n$ targets. The proton layer is located only at the back surface of the first target. The remaining targets are devoid of any contaminant hydrogen-rich materials. 

In the two-stage setup, the proton layer gets accelerated by the electrostatic field developed through the interaction of the first laser sub-pulse with the first substrate. The second laser sub-pulse in the mean time, travels to the second target, interacts with it and sets up a longitudinal electric field. The travelling proton layer passes through the second substrate and gets an extra boost from this electric field. If the arrival time for the second laser sub-pulse at the second target is properly adjusted, the proton layer gets an appreciable energy increase. In our numerical simulations we implement the time delay by performing several pre-simulation runs from which we determine the optimum timing of the laser pulse on every subsequent target. It is interesting to note that the dependence of the final proton energy (and also the laser-to-proton energy transfer efficiency) on time delays exhibits a resonant behavior, so that any deviation from the most optimal distribution will lead to lower final proton energy. Therefore, the delay times will have to be properly adjusted in an actual experiment until the highest proton energy is achieved. Moreover, each individual interaction stage will introduce additional transverse divergence in the proton beam, which may lead to particle loss. The beam divergence is due to the generated transverse electric field (induced by the laser ponderomotive force gradient in the transverse direction). In order to limit this effect, one needs to assure that the laser pulse width is larger than the transverse size of the proton layer (e.g. by using layered microstructures as done in reference~\cite{schwoerer06}).                

\begin{figure}
\centerline{\includegraphics[angle=-90,width=1.\columnwidth]{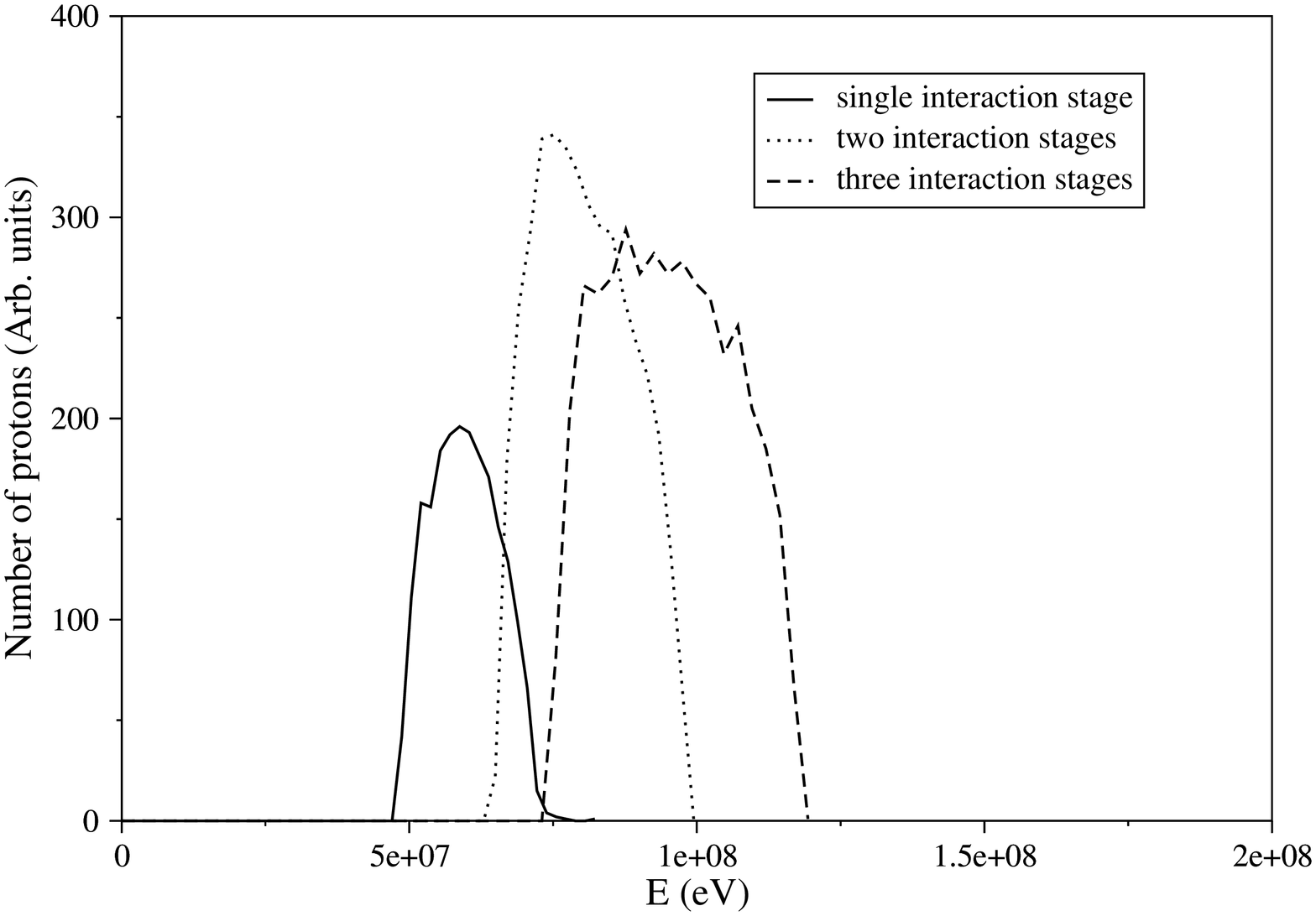}} 
\caption{
Proton energy distributions for three different interaction stages. Solid line represents a single interaction stage (no laser splitting), dotted line represents a double interaction stage (single laser splitting) and dashed line represents a tripple interaction stage (double laser splitting).}\label{fig2}
\end{figure}

The results of PIC simulations show that with the two-stage splitting the final average energy of the accelerated protons reaches $E^{(2)}_p$=81.5 MeV, as opposed to $E^{(1)}_p$=60.5 MeV (where the superscript denotes the number of interaction stages) for the  conventional single target assembly, which is an increase of $\sim$ $35 \%$. Using the procedure described above, we have also simulated a 3-stage interaction scheme, in which case the main laser pulse is split into three sub-pulses of equal intensity $I=I_0/3$ that are made to interact with three targets with the same physical parameters described above. The final average proton energy in this 3-stage setting reaches $E^{(3)}_{p}$= 96.5 MeV, which is $\sim$ $60 \%$ energy increase as compared to the single interaction case or $\sim$ $19 \%$ as compared to the 2-stage procedure. Figure~\ref{fig2} also shows the proton energy distributions for the three interaction stages. Gradual increase in the peak proton energy is clearly seen. 

As the number of splits $n$ increases one should expect a gradual increase in the final proton energy. Of course there is a limit on the number of interaction stages, which would yield higher proton energies. This limit is related to the fact that the intensity of laser sub-pulses has to be high enough so that the laser ponderomotive force can still push electrons out of the target, thus setting up an accelerating electric field for protons. For estimation purposes, the number of splits $n\sim a_0^2$, where $a_0=eE/(mc\omega)$ is the laser relativistic parameter. In order to study how the proton energy changes with greater number of splits ($n>3$) as well as splitting ratio between laser sub-pulses, we have used earlier developed model~\cite{velchev07} for the longitudinal electric field, briefly described in the next paragraph. 

\begin{figure}
\centerline{\includegraphics[width=1.\columnwidth]{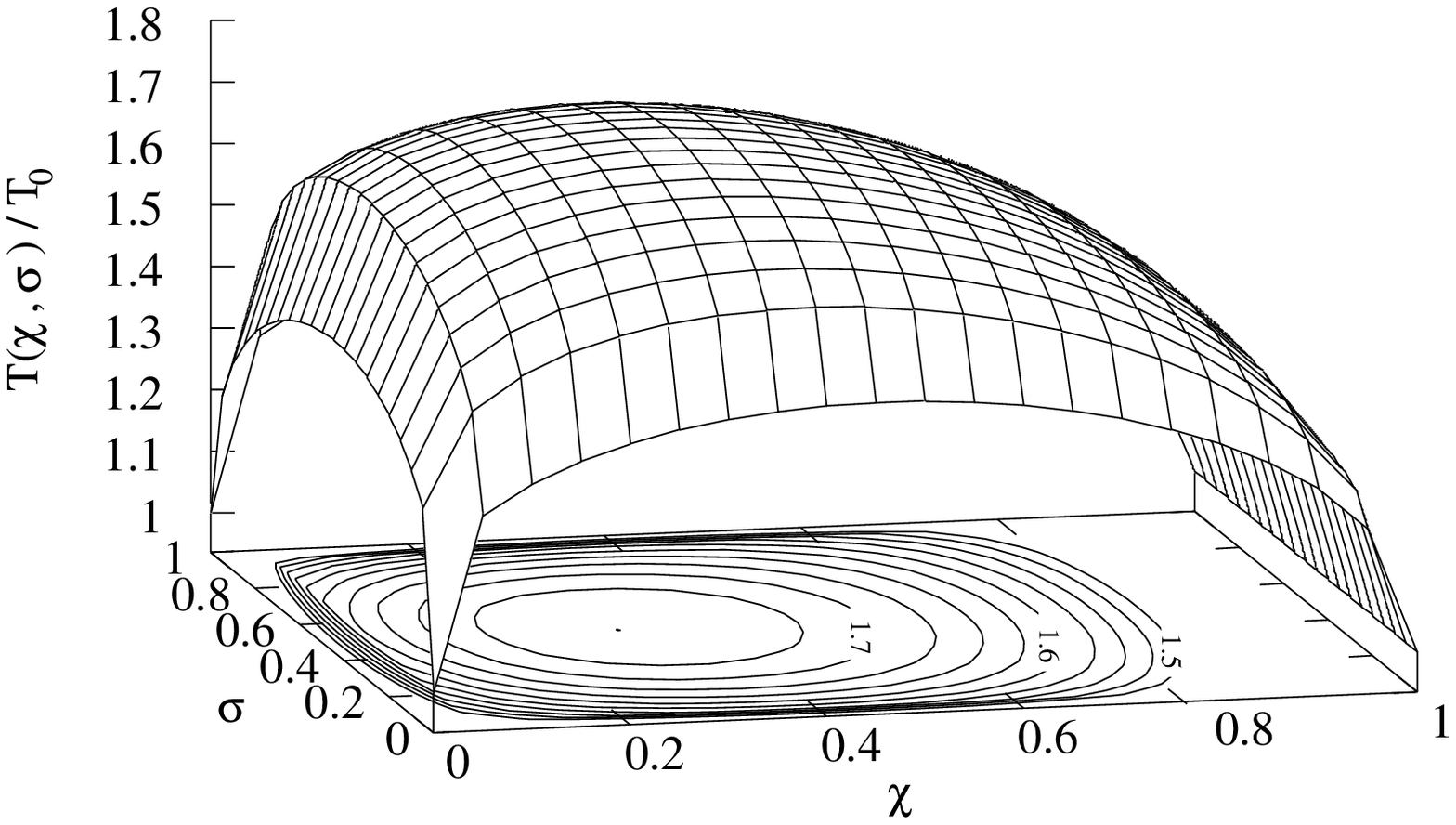}}
\caption{ Final proton energy as a function of splitting ratios $\chi$ and $\sigma$ }\label{fig3}
\end{figure}

The model is based on the fact that the accelerating electric field can be approximated by that of a charged cylinder of radius $a$ and thickness $2r_0$, having the following mathematical form (on the cylinder's axis $x$),
\begin{eqnarray}
&&E(x,t) = \frac{k Q_0\eta(t-\frac{|x|}{c}) }{a^2 r_0} \Bigg[ %
\sqrt{(x-r_0)^2+a^2}-\sqrt{(x+r_0)^2+a^2} \nonumber \\
&&+2x\left\{%
\begin{array}{ll} 
1   & ,\ |x| \leq r_0 \\
r_0/|x| & ,\ |x| > r_0 
\end{array} \right. \Bigg], \label{E3D_x_t}
\end{eqnarray} 
where $Q_0$ is the charge of the target if all electrons are expelled, $k=1/(4\pi \epsilon_0)$ and
$\eta(t)$ is the proportion of the expelled electrons as a function of time that can be approximated by the following expression\cite{velchev07},
\begin{equation}
\label{eta_t}
\eta(t) = \gamma \left\{ %
\begin{array}{ll}
e^{-\alpha(t-t_0)^2} & ,\ t \leq t_0 \\
\delta + (1-\delta) e^{-\beta(t-t_0)} & ,\ t > t_0
\end{array} \right. ,
\end{equation}
where $\gamma$ is the fraction of the electrons expelled at the peak of the laser pulse, $\delta$ is the fraction of the initially expelled electrons that never return to the target, $t_0$ is the arrival time of the peak of the laser pulse at the target, $\alpha=4 \ln{2} / \tau^2$ is a constant that depends on the pulse width (FWHM) used in the PIC simulation, $\beta$ is the rate of return of the expelled electrons. These numerical factors are functions of laser intensity and have been tabulated using the PIC simulations.  The equation of motion for a proton interacting with the field distribution (\ref{E3D_x_t}) is:
\begin{eqnarray}
\label{eq_mo}
&&\frac{d}{dt}\left(\frac{v}{\sqrt{1-v^2/c^2}}\right)= \frac{e}{m_p} E(x,t), \nonumber \\
&&\frac{d}{dt}x(t)=v(t),
\end{eqnarray}
where $m_p$ is the proton mass, and $e$ is the elementary charge. Eqs.~(\ref{eq_mo}) have been solved numerically for a wide range of splitting ratios $\chi=I_1/I_0$ ($I_1$ is the intensity of the first laser sub-pulse) and $\sigma=I_2/((1-\chi)I_0)$ ($I_2$ is the intensity of the second laser sub-pulse) in the tree-stage interaction scheme. Figure~\ref{fig3} shows the final average proton energy as a function of the splitting ratio parameters normalized to the proton energy obtained from a single interaction stage. As one can see, the maximum in the proton energy occurs when $\chi=1/3$ and $\sigma$=1/2 (corresponding to three laser sub-pulses with equal intensities $I_0/3$). Any other combination of splitting ratios leads to lower final proton energy. It should be noted that the two-stage results are recovered from this figure when one of the splitting parameters ($\chi$, $\sigma$) are equal to 0 or 1. In this case, the maximum in proton energy is reached at exactly equal splitting of the laser pulse into two sub-pulses with intensity $I_0/2$. The fact that equal pulse splitting leads to maximum proton energy is not a mere coincidence and its origin will be discussed in the next section.                                                        
Using Eqs.~\ref{eq_mo}, we have also investigated how the final proton energy depends on the number of amplification stages, shown in figure~\ref{fig4}. As one can see, the final proton energy increases as the number of spitting stages becomes larger. 

\section{Interpretation of obtained results}     
Results obtained in the previous section pose two interesting questions. Firstly, why splitting of the single interaction into multiple sub-stages gradually increases the amount of energy transferred from the laser pulse to protons? And secondly, why only \emph {equal} laser pulse splitting leads to maximum proton energy? One possible answer to the first question can be obtained from the analysis of the microscopic dynamics of protons in the electric fields created through the interaction of each laser sub-pulse with each target. The analysis shows that the splitting procedure allows for more efficient utilization of the energy of the longitudinal electric fields in case of multiple interaction stages as opposed to that in the single, which gradually leads to higher proton energy. However, the same analysis does not yield a satisfactory answer to the second question. As mentioned in the introduction, proton acceleration by high power lasers can be qualitatively viewed from a thermodynamic perspective as an energy exchange process between hot (with initial temperature $T_h$) and cold (with initial temperature $T_c$) objects. Even though on a microscopic level the interaction potentials and the equations of motion for atoms on one hand and electrons and ions on the other are different, on a macroscopic level these two problems are quite similar in that they deal with energy flow and can be described using thermodynamic approach. The problem at hand may be formulated as follows: what is the most efficient method of exchanging the energy between the hot and cold objects, so that in the end the initially hot object becomes cold and initially cold object becomes hot? The most simple way by which one can exchange the heat between both objects is through putting them into the thermal contact with each other, so that in the end their final temperature is just half of the sum of their initial temperatures. The entropy change for this particular process corresponds to a maximum in the entropy gain, making it completely irreversible and least efficient in the sense of energy exchange between both objects. From a thermodynamic point of view~\cite{lifshits}, the efficiency of the energy transfer from the hot object to the cold is at a maximum for those processes for which the entropy change tends to zero. Therefore, the problem is reduced to finding those processes that minimize the entropy gain. 

At this point, we would like to refer the reader to the Appendix section where we introduce the so-called "perfect heat exchange" problem~\cite{mishchenko} and show that a very efficient heat exchange process between the hot and cold objects may be established using a splitting procedure. It is shown that just like in the present problem of the energy exchange between the laser pulse and protons, the splitting procedure in the "perfect heat exchange" problem also leads to more efficient energy exchange between the hot and cold objects. It is also confirmed that the increased energy transfer efficiency is accompanied by decreasing increment in the total entropy change. This leads to the formulation of the following conjecture: the increased energy transfer efficiency between the laser pulse and protons in the multistage model is due to the reduction in the total entropy increment for the laser-target system.         

In addition, one can also arrive at the above conjecture with the help of PIC simulations. The total entropy change for the laser/target system can be estimated using the following expression,
\begin{equation}
\Delta S=\Delta S_{l}+\Delta S_{e}=\Delta E\left(1/T_{e}-1/T_{l}\right),
\end{equation}             
where $\Delta E$ is the energy transferred from the laser pulse to the target and $T_{e}$ is the initial temperature of plasma target (usually in the several keV range). Since the effective laser temperature $T_{l}\gg T_{e}$ (see below), one can neglect the term $1/T_{l}$ in the brackets and for the multiple splitting case one obtains,
\begin{equation}
\Delta S_n= \frac{1}{n}\sum_{i=1}^n\Delta E_n/T_{e},
\end{equation}    
where $\Delta E_n$ is the energy transferred from the laser sub-pulse of intensity $I_0/n$ to the $n^{th}$ target. Using the PIC simulation results, we have calculated the total energy transferred from the laser pulse to the target (electrons, ions and protons), as a function of the laser splitting ratio $I/I_0$ shown in Figure~\ref{fig5}\,a. As one can see, as the laser intensity becomes smaller, the amount of energy $\Delta E_n$ transferred to the target is reduced, which eventually leads to a smaller increment in the entropy change for the total system. It is an expected result since the entropy production rate depends on the induced (by the laser pulse) temperature, density, pressure gradients in a plasma slab. Lower laser pulse intensity induces smaller hydrodynamic gradients and lower heat fluxes, leading to smaller entropy production rates and correspondingly smaller total (particles and fields) entropy change. At the same time, Figure~\ref{fig5}\,b shows the efficiency of energy transfer from the laser pulse to protons as a function of laser energy obtained from the PIC simulations. As one can see, the energy transfer efficiency increases with decreasing amount of energy in the incoming laser pulse, confirming the earlier made conjecture concerning the relation between the energy transfer efficiency and the entropy change for the laser-target system.  

Finally, we would like to discuss the applicability of the thermodynamic interpretation of the multistage laser-proton acceleration. Expressions for the final temperatures or stored energies presented above have been derived under the assumption that both systems reach a state of thermodynamic equilibrium after the thermal contact has been established. In a case of laser interacting with the target, the thermal equilibrium between the laser and target is never achieved. Therefore, expression~(\ref{tc}) cannot be directly used to evaluate the amount of energy stored in the proton beam. However, the fact that the thermal equilibrium between the laser and target is never achieved only results in lower value for the total amount of energy transferred between both systems, without changing the main conclusion that the efficiency of energy transfer from the laser pulse to protons increases with increasing number of splits. In addition, unlike in the "perfect heat exchange" problem where both the hot and cold reservoirs have been in their own corresponding equilibrium states (with  well defined temperatures, entropies, etc.) prior to the thermal contact (the total closed system incorporating the hot and cold reservoirs however is a non-equilibrium system), in the laser-target system the laser pulse before the interaction with the target is not in a state of thermodynamic equilibrium. This fact however should not affect the main conjecture of the paper for the reasons presented below. 

In order to make a connection between the results of statistical mechanics as applied to the problem of heat exchange between hot and cold reservoirs on one hand and laser-target interaction on the other, definitions for the thermodynamic quantities such as temperature and entropy for a laser system are needed. One may also assume that the target substrate is in the state of thermodynamic equilibrium before and shortly after the interaction with the laser pulse, since the energy relaxation time $\tau_{ee}$ (or Maxwellization) between the hot and cold electron populations of the substrate is on the order of $\sim 10^{-14}$ s~\cite{eliezer} for electron temperature of 1 keV, which is much shorter than the proton acceleration time (even though the relaxation time for the ion component $\tau_{ii}=\tau_{ee}\sqrt{\frac{m_i}{m_e}}$ and that between electrons and ions $\tau_{ei}=\tau_{ee}\frac{m_i}{m_e}$ are longer than $\tau_{ee}$, one may assume Maxwellian distribution for ions with its own temperature unequal to that for the electrons). The proton layer distribution function throughout the acceleration process may be assumed shifted Maxwellian with average velocity equal to that acquired by protons in the accelerating electric field (once the particles left effective acceleration region). Since the incoming laser light is not a black body radiation and is not in a thermodynamic equilibrium (in a sense that its entropy is not at a maximum), its effective temperature has certain ambiguity in its definition and still remains a matter of debate~\cite{essex03}. Nonetheless, the thermodynamic definition of temperature given by expression,
\begin{equation}
\label{entr}
\frac{1}{T}=\left(\frac{\partial S}{\partial E}\right)_V,
\end{equation}         
where $S$ is the entropy of the laser pulse and $E$ is its total energy, leads to the so called brightness temperature of the laser light~\cite{mungan05}. The entropy of the laser light is usually calculated by counting states for identical bosons~\cite{lifshits,essex03}. In this case the entropy is due to the randomness of the phase of each field quantum. Using expression~(\ref{entr}) one can find the brightness temperature for laser light, given by the following~\cite{lifshits,essex03},
\begin{equation}
\label{temp}
k_BT=\frac{\hbar \omega}{\ln\left(1+1/\bar{n}\right)},
\end{equation}
where $\bar{n}$ is the mean photon occupation number, which is related to the laser specific intensity $I_\omega $ (laser intensity per unit frequency) through the relation $\bar{n}=4\pi^3 c^2I_\omega /\hbar \omega^3$. Even for low intensity laser light, the mean occupation number is very large, so that the limit $\bar{n}>>1$ can be easily applied to the ultra-high intensity lasers. In this case, a very simple relation between the laser temperature (for mode $\omega$) and its spectral intensity is obtained,
\begin{equation}
\label{temp1}
k_BT_\omega=\pi \lambda^2 I_\omega.
\end{equation}
This shows that each individual mode $\omega$ has its own temperature $T_\omega$. Note that for laser intensity $I_0$, carrying frequency $\omega_0$ and spectral width $\Delta \omega$ used in the PIC simulations, the laser temperature (at the central frequency $\omega_0$) reaches the staggering value of $\approx 10^{21}$ $K$. Thus, for any practical applications one can assume that the ultra-high intensity laser light is a system with infinite temperature. In this respect, the laser light can be considered to be a collection of non-interacting (because of linearity of electrodynamic equations) stationary (see below) sub-systems, each with its own temperature $T_{\omega}$~\cite{lifshits}. Since different frequency modes in the pulse do not interact with each other, the relaxation time for the propagating laser light into the equilibrium state is effectively infinite in the absence of matter. This in turn implies that a propagating laser pulse in itself may effectively be considered a static system in a state that resembles an equilibrium distribution (even though initially it was created by the system far away from the equilibrium and its spectral and angular distributions do not resemble the "black-body" radiation spectra) since its distribution function does not change in time as it propagates to and from the target. In addition, one may assume (neglecting the effects of the change in the laser angular and spectral distributions during the course of interaction) that the entropy change in the laser pulse as a result of its interaction with the target is smaller (in its absolute value) than that due to the target system (since laser effective temperature is infinite), so that the total entropy change is solely due to the target. As was mentioned in the introduction section, shorter relaxation times in the system lead to lower entropy increase and correspondingly higher energy transfer efficiency. Since the relaxation time for the electron component is equal the inverse of the electron-electron collision frequency $\tau_{ee}=\nu_{ee}^{-1}(v_{Th})$, higher laser intensity leads to higher final electron temperature and correspondingly longer relaxation times and lower energy transfer efficiency (this argument also confirms the main conjecture of this work albeit from different consideration). In this respect, the arguments presented above give qualitative justification to the thermodynamic analogy between the "perfect heat exchange" problem on one hand and the multi-stage laser-plasma interaction model on the other.            

The main conclusion that one can draw from the example above is that splitting of the single interaction stage into multiple sub-stages is an effective way of reducing an irreversible component in the total interaction cycle no matter how this interaction looks like (laser-matter or matter-matter), thus increasing the effectiveness of the "pump". This is why the splitting procedure should also lead to higher proton energies in the laser-matter interaction experiments, since it increases the effectiveness of the energy transfer from the laser pulse to protons. 

\section{Discussions and Conclusion}
The problem of proton acceleration by high-power lasers has been revisited in this work using 2D PIC simulations as well as an analytical 3D model. It was shown that significant energy gain in the final proton energy is possible if one introduces a multistage interaction scheme as opposed to a conventional single laser/target interaction setup. Many recent investigations concerning the proton acceleration looked at the kinematic/dynamic aspect of this problem, specifically the underlying physics behind the particle acceleration. However, the problem of how to increase the energy transfer efficiency from the laser to accelerated particles has not been addressed to the same degree of scrupulousness. As shown in the present work, the multistage interaction model offers significant gain in the efficiency of energy transfer from the laser to accelerated particles. A thermodynamic analogy has been offered to elucidate this effect. According to the model, the splitting of a single interaction site into multiple stages is an effective way of reducing an irreversible component in the energy exchange process between the laser and protons. As a result more laser energy is transformed into proton kinetic energy. It was shown that in a three-stage setting, there is $\approx 60 \%$ increase in the energy efficiency of the laser accelerator as compared to a single interaction scheme. At the same time according to the results of our 3D model, it should be possible to increase the energy efficiency by more than 100$\%$ for a six-stage interaction setting without using more powerful lasers. Based on these results we conclude that the multi-staging procedure represents a step forward in increasing the energy efficiency of laser-ion accelerators with the potential of achieving significant increase in the final proton energies suitable for practical applications.        

\appendix
\section{Perfect heat exchange problem and its relation to laser-matter interaction}
In what follows, we describe the "perfect heat exchange" problem and discuss its relation to laser-to-proton energy transfer efficiency. Let us suppose that we have hot and cold reservoirs (for a sake of simplicity we shall assume that both objects have the same size and mass and consist of an ideal gas) with initial temperatures $T_h$ and $T_c$ correspondingly. The task is to find such a heat exchange process that would maximize the energy exchange between both reservoirs, so that in the end the hot reservoir becomes cold and the cold reservoir becomes hot. 
    
If initially hot and cold reservoirs are split into $n$ equal pieces each and subsequently every individual hot piece is put into thermal contact with each individual cold piece (without mixing them) in a sequential manner as shown in figure~\ref{fig6}, then using the thermal balance equations, it can be easily shown that the final temperature of initially hot/cold objects (formed by putting back together initially hot/cold pieces to form new cold/hot objects) can be smaller/greater than $(T_h+T_c)/2$. The general expression for the final temperature of initially hot/cold reservoirs for $n$ equal splittings is given by the following expressions~\cite{mishchenko}:
\begin{eqnarray}
\label{temp0}
&&T^{(fin)}_{h} =
\frac{\sum_{i=1}^{n}\left(\delta_{i,n}T_c+\delta_{n,i}T_h\right)}{n}, \\
\nonumber
&&T^{(fin)}_{c} =  
\frac{\sum_{i=1}^{n}\left(\delta_{i,n}T_h+\delta_{n,i}T_c\right)}{n}, \\ \nonumber 
&&\delta_{i,j}=\frac{1}{2^{i+j}(i-1)!}\sum_{k=1}^{j}2^k\frac{(i+j-k-1)!}{(j-k)!} 
\end{eqnarray}                                   
It can be shown that in the limit $n\rightarrow \infty$, $T^{(fin)}_{h}=T_c$ and $T^{(fin)}_{c}=T_h$ and a perfect heat exchange process between hot and cold objects is established. Assuming that both objects are an ideal gas, the entropy change for the process involving $n$ equal splittings has the following form:
\begin{equation}
\frac{\Delta S}{C_p}=\ln\left[\frac{\prod_{i=1}^n\left(\delta_{i,n}T_h+\delta_{n,i}T_c\right)\left(\delta_{i,n}T_c+\delta_{n,i}T_h\right)}
{T_h^nT_c^n}\right]/n
\end{equation} 
where $C_p$ is the specific heat capacity of the material. As the number of splits $n$  increases, the entropy increment $\Delta S$ decreases (remaining positive) and in the limit $n\rightarrow \infty$, the entropy change $\Delta S\rightarrow 0$. We would like to note that even though we used an ideal gas in the calculation of the entropy change, the same conclusion can be drawn if one were to use any other system.  
   
At this point we would like to draw the reader's attention to the fact that the multistage laser-target interaction scheme may be qualitatively viewed as a particular case of the "perfect heat exchange" problem as presented above. In the proposed multistage model, a given laser sub-pulse interacts with a given substrate only once. However, the proton layer gets $n$ energy kicks originating from the electric fields created by the interaction of the $i$-th laser sub-pulse with the $i$th substrate for $i \in (1,n)$. In this respect the proposed multistage model may be viewed as a "heat exchange" problem between a small cold object of mass $m$ and initial temperature $T_c$ (protons) and a large hot reservoir of mass $M$ and initial temperature $T_h$ (laser pulse), split into $n$ small objects (sub-pulses) each having mass $M/n$. 

Moreover, in the derivation of eqs.~\ref{temp0} it is assumed that the interface between the hot and cold reservoirs is an ideal thermal conductor, so that there is no heat loss (e.g. radiative losses) during the energy flow. In the laser-target interaction scheme the role of the interface is taken by the substrate material (electrons and ions of the substrate), which of course is not an ideal thermal conductor. This results in reduced energy gain for protons, while the splitting scheme remains beneficial. In addition, when the intensity of the laser pulse drops below a certain threshold value (so that electrons of the medium oscillate in phase with the incident light), the energy absorbed by the electrons of the substrate (through the absorption of the laser photons) gets reemitted back into the vacuum, so that in the end there is almost no net laser energy deposited in the target (classical skin effect). The threshold value for laser intensity at which this occurs is $a_0\approx 1$. When the laser sub-pulse's amplitude drops below this value, there is virtually no net energy flow from the laser to the substrate and subsequently to the protons. Therefore, unlike the perfect heat exchange problem where the energy flow between the hot and cold reservoirs is always present no matter what the splitting ratio $n$ is, in a case of laser-target interactions it would not be beneficial to split the laser pulse into too many sub-pulses. The threshold for the laser amplitude defined earlier determines the actual number of splitting stages for which one should expect energy flow from the laser to protons.
      
Using the thermal balance equations one can readily arrive at the expression  for the final amount of energy stored in initially cold (with internal energy $E_c$) and hot (with internal energy $E_h$) objects as well as the corresponding entropy change,
\begin{eqnarray}
\label{heat_pr}
&&E^{(fin)}_{c}=\frac{E_c}{\left(1+\frac{1}{n\delta}\right)^n}+ E_h\delta \left(1-\frac{1}{\left(1+\frac{1}{n\delta}\right)^n}\right) , \label{tc} \\  \nonumber
&&E^{(fin)}_{h}=E_h-E_h\delta \left(1-\frac{1}{\left(1+\frac{1}{n\delta}\right)^n}\right) 
+E_c \left(1-\frac{1}{\left(1+\frac{1}{n\delta}\right)^n}\right) \\ \nonumber
&&\frac{\Delta S}{C_p} =\delta \ln \frac{E^{(fin)}_{c}}{E_c}+ 
\frac{1}{n}\ln\frac{\prod_{i=1}^n\left[E_h\left(1-\frac{1}{\left(1+\frac{1}{n\delta}\right)^i}\right)+
\frac{E_c/\delta}{\left(1+\frac{1}{n\delta}\right)^i}\right] }{E^N_h} , \label{Se} 
\end{eqnarray}        
where $\delta=m/M$. For a given value of $\delta<<1$, the final energy stored in a small object $E^{(fin)}_c$ increases with the number of splits $n$, quickly approaching the limiting value $E_h\delta$. The total entropy change also decreases with the number of splits as in the general case discussed earlier, however unlike in the "perfect heat exchange" problem, it never goes to zero (in the limit $n \rightarrow \infty$) but approaches a finite value, which depends on $\delta$. This indicates that there is a limit on the efficiency of the energy transfer between the two objects in this particular case, leading us to conclusion that there may also be a limit on how much of the laser energy can be transferred to protons in the multistage laser-matter interaction scheme. This limit can also be seen in Figure~\ref{fig4} where the final proton energy gradually approaches saturation with the number of interaction stages. The same behavior was also observed in the PIC simulation results. In addition, the PIC simulation also showed that only equal laser pulse splitting led to the highest proton energy. Any other combination of splitting parameters would lead to lower proton energy. As we mentioned earlier, this peculiarity is not coincidental. Analyzing the thermal balance equations in the case of arbitrary distribution of the masses of hot small objects (energies of laser sub-pulses), it can be shown using the variational principle that only homogeneous distribution (equal splitting) will lead to the maximum in the final temperature (energy of protons) of initially cold object.                     

\begin{acknowledgments} 
This work is in part supported by NIH grant No. CA78331d, Strawbridge
Family Foundation, and Varian medical systems. E. F. is thankful to E. Mishchenko and P. Pshenichka for introducing us to the perfect heat exchange problem. E. F. would also like to thank A. Smolyakov for stimulating discussions.     
\end{acknowledgments}



\begin{figure}
\centerline{\includegraphics[width=1.0\columnwidth]{fig4p}}
\caption{ 
Final proton energy as a function of the number of amplification stages}\label{fig4}
\end{figure}

\begin{figure}
\centerline{\includegraphics[width=1.\columnwidth]{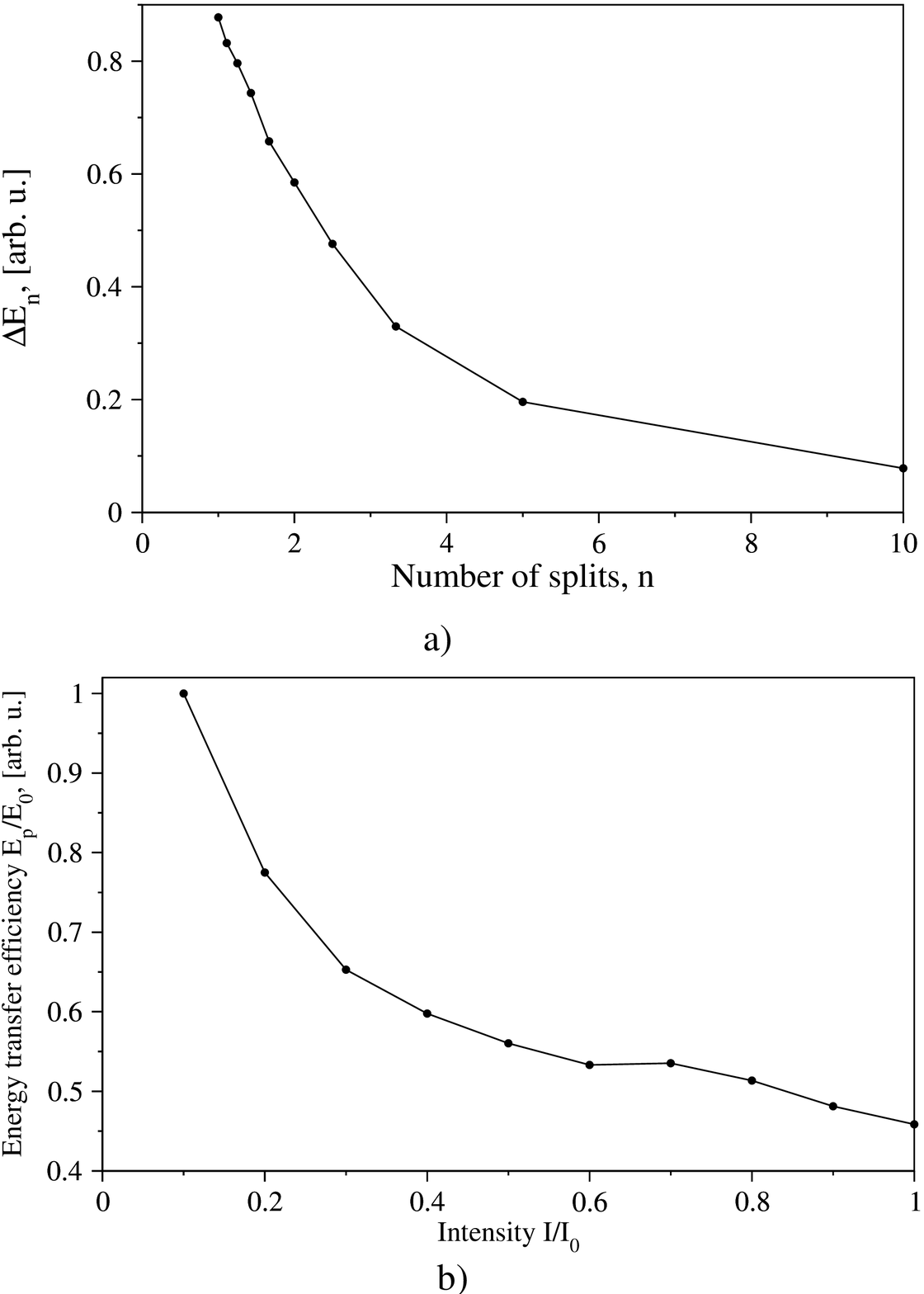}}
\caption{a) Energy transferred from the laser sub-pulse to the target versus the laser splitting ratio $n=I_0/I$. b) Normalized energy transfer efficiency from laser pulse to protons versus normalized laser intensity. $I_0=1.9 \times 10^{21}$ W/cm$^2$}\label{fig5}
\end{figure}

\begin{figure}
\centerline{\includegraphics[width=1.\columnwidth]{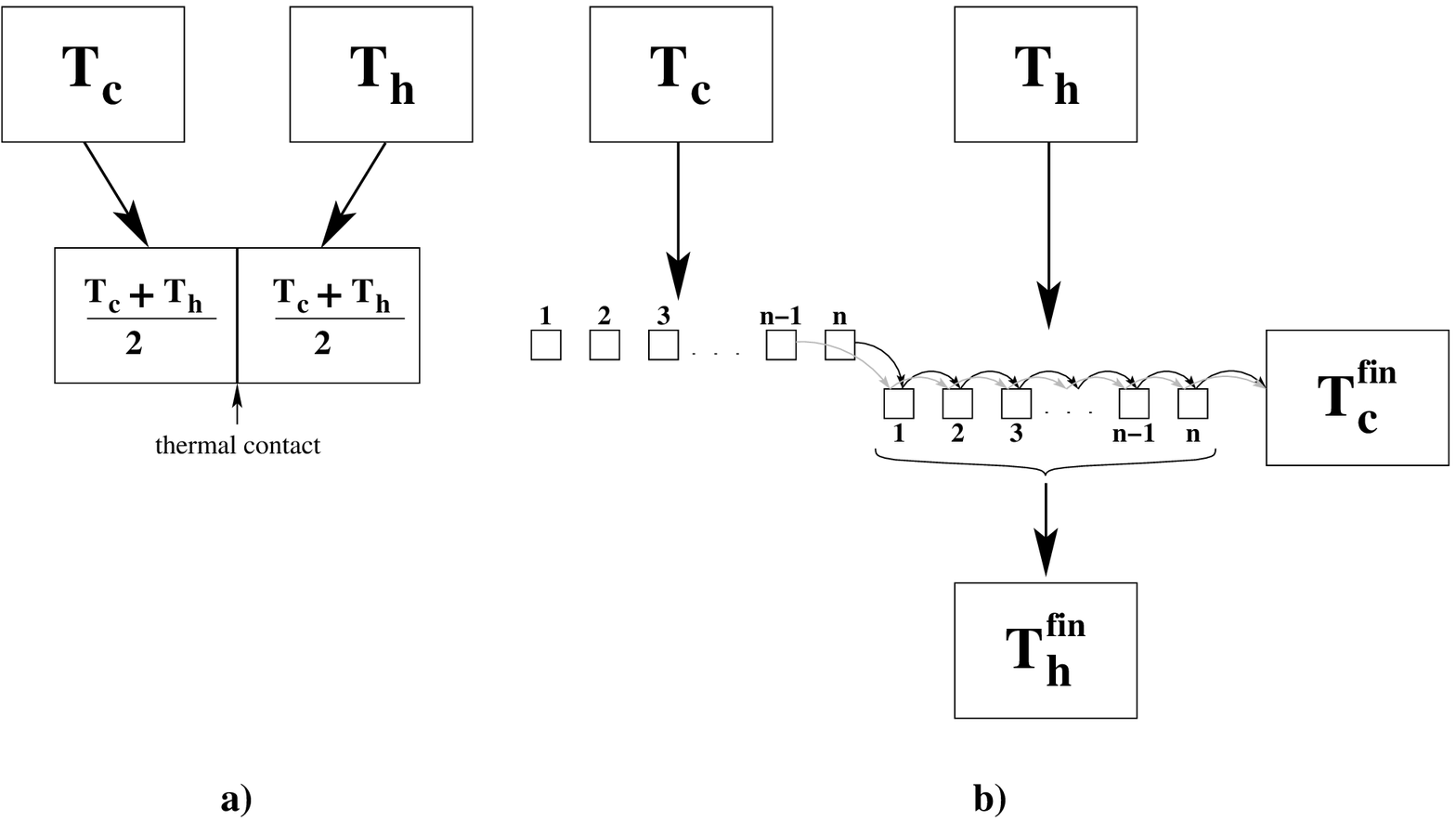}}
\caption{ 
Schematic diagram for two heat exchange processes. a) The hot and cold reservoirs are put into thermal contact with each other leading to temperature equalization. The entropy gain is maximal for this process.  b) The hot and cold reservoirs are split into $n$ pieces each that are put into thermal contact with each other in a sequential manner. The limit $n \rightarrow \infty$, corresponds to totaly reversible heat eachange process with zero entropy gain.}\label{fig6}
\end{figure}

\end{document}